\documentclass[aps,prl,floatfix,twocolumn,superscriptaddress]{revtex4-1}
\usepackage{amsmath,amssymb}
\usepackage[utf8x]{inputenc}
\usepackage{graphicx}
\usepackage{subfig}
\usepackage{color}
\def\<{\langle}
\def\>{\rangle}
\def\(({\left(}
\def\)){\right)}
\def\[[{\left[}
\def\]]{\right]}

\begin{document}

\title{The Super-Potts glass: a new disordered model for glass-forming liquids}
\author{Maria Chiara Angelini, Giulio Biroli}
\affiliation{IPhT, CEA/DSM-CNRS/URA 2306, CEA Saclay, F-91191 Gif-sur-Yvette Cedex, France}
\begin{abstract}
We introduce a new disordered system, the Super-Potts model, which is a more frustrated 
version of the Potts glass. Its elementary degrees of freedom are variables that can take $M$ values
and are coupled via pair-wise interactions.   
Its exact solution on a completely connected lattice demonstrates that for large enough $M$ it belongs
to the class of mean-field systems solved by a one step replica symmetry breaking Ansatz. 
Numerical simulations by the parallel tempering technique show that in three dimensions it displays 
a phenomenological behaviour similar to the one of glass-forming liquids. The Super-Potts glass is therefore the long-sought disordered model allowing one to perform extensive and detailed studies of the Random First Order Transition in finite dimensions. 
We also discuss its behaviour for small values of $M$, which is similar to the one of spin-glasses in a field. 
\end{abstract}

\maketitle


Glass forming liquids have a very peculiar and rich phenomenology \cite{BBRMP}. Dynamical correlation 
functions are characterized by a two-steps relaxation indicating that a finite fraction of degrees of freedom, {\it e.g.}
density fluctuations, takes a longer and longer time $\tau$ to relax. This time-scale actually grows very rapidly---more than 
14 orders of magnitude in a rather restricted window of temperatures---and can be fitted by the Vogel-Fulcher-Tamman
law, hence suggesting a possible divergence at finite temperature. The slowing down of the dynamics is accompanied by the growing 
of dynamical correlations, which can be measured by a four point susceptibility. This function 
displays at time $\tau$ a peak, that grows decreasing the temperature and is related to the number of molecules 
that have to move in a correlated way in order to make the liquid flow. \\
One of the most influential results obtained in the field of the glass transition was the discovery by Kirkpatrick, Thirumalai and
Wolynes \cite{RFOT} that some---apparently unrelated---fully connected Mean-Field (MF) disordered systems, like the Potts glass, display a phenomenology
very similar to the one described above. This set the stage for an approach to the glass transition problem that combined disordered systems,
Mode-Coupling and Adam-Gibbs theories and culminated in the development of the Random First Order Transition (RFOT) theory \cite{RFOTbook}.
Although structural liquids do not explicitly contain quenched disorder in the Hamiltonian, they are frustrated and characterized
by a very complicated rugged energy landscape. This is the key element they have in common with several disordered
systems and that is at the origin of the relationship cited above. 
MF disordered systems divide in two classes: some have a phenomenology similar to glass-forming liquids, 
others to spin-glasses. 
The former are the ones for which, in replica language, the one step replica symmetry breaking (1RSB)
approximation is exact \cite{1RSB}. For these models the relaxation time is known to diverge at a finite temperature, 
called $T_d$ \cite{pspin_dyn}. This transition was shown to be identical to the one predicted by the Mode Coupling theory of the glass transition \cite{BBRMP}.
Below $T_d$ ergodicity is broken. The phase space is fractured 
into a number of states $\mathcal{N}$ 
that is exponential with the size $N$ of the system: $\mathcal{N}\propto e^{N\Sigma}$ ($\Sigma$ is called complexity or configurational entropy).
The system undergoes a thermodynamic phase transition \`a la Kauzmann 
at a smaller temperature $T_K<T_d$, where the configurational entropy vanishes and hence number of states that dominate the Boltzmann measure becomes sub-exponential \cite{Cavagna}. The order parameter for this transition is the overlap $q$ measuring the similarity between two different replicas of the system (characterized by the same realization of the disorder). Its distribution, $P(q)$, shows a single peak at $q_{RS}$ for $T>T_K$ and two distinct peaks $q_0$ and $q_1$ for
$T<T_K$. The lowest value, $q_0$, corresponds to the two replicas being in configurations 
belonging to two different amorphous states, whereas 
the higher one, $q_1$,  to configurations belonging to the same state.
There is however another class of MF disordered systems, the spin-glasses, 
characterized by a quite different behavior. 
They display a continuous transition and are solved by  the Full Replica Symmetry Breaking (FRSB) Ansatz \cite{FRSB}. 
Dynamical correlation functions do not show any two-step relaxation, the four point susceptibility is not peaked,
$P(q)$ has a continuous support below the transition and $T_K=T_d$.\\
In view of the forementioned analogy between structural glasses and MF 1RSB disordered models 
and of its relevance for RFOT theory, the numerical results on finite dimensional counterpart of MF 1RSB 
systems were deceiving. It was found that the usual fate of these systems, once studied on finite dimensional lattices,
is to display either a continuous spin glass transition or no transition at all!
For instance, the MF Potts glass \cite{Potts}, the model from which RFOT theory originated, 
is characterized by a glass transition for any $p>4$, 
where $p$ is the number of values that Potts variables can take, 
but in three dimension it does not show any transition for $p=10$ \cite{Potts3D}. 
The problem of the disappearing of the 1RSB phenomenology in finite dimension could be a signal
of the fragility of the 1RSB theory out of MF, and poses the question of the validity of RFOT 
in $D=3$ as discussed in a series of paper by Moore and collaborators \cite{moore}. 
In a recent work \cite{NO1RSBbethe} it was pointed out that the MF disordered models 
studied so far are not frustrated enough and even simple local fluctuations are enough to change their physics 
(see also \cite{eastwood02}). 
This is well illustrated by their change of behaviour on Bethe lattices, 
which provide a better mean-field like approximation than fully-connected models
since have finite connectivity and, hence, allow one to take into account the kind 
of local fluctuations present in finite dimensions.  
One should not conclude however that there are not models or results connecting MF theory to the behavior of 
finite dimensional glass-forming liquids. Indeed, there are. Lattice glass models display the correct 
phenomenological behavior and they belong to the 1RSB class when solved on a Bethe lattice \cite{LG1,LG2}.  
A particular form of a disordered 5-spin model appears to behave correctly too \cite{5spin3D_KZ}. 
Finally, hard spheres in the limit of infinite dimensions do display a 1RSB transition \cite{HS_highD}.
However, from the point of view of the quest of finding simple finite dimensional models displaying a glass transition,  
all these systems suffer from one or more limitations: they are either too hard to simulate in finite dimensions or 
they display a crystal phase that preempts the existence of the glass transition and deep super-cooling or they 
do not have pair-wise interactions, which makes them difficult to be analyzed in finite dimension, in particular by real 
space renormalization group methods. \\
The aim of this work is to introduce and study a model that short-circuits these problems and therefore offers a 
new way to test RFOT theory and to answer questions on glassy physics. We call it the {\it Super-Potts model}. 
It is similar to the modifications of the Potts glass introduced and studied in \cite{PermutationPotts,PermutationPotts2}, which display a continuous transition and not the discontinuous one that we are looking for. 
Its degrees of freedom are variables that take $M$ values, as in the usual Potts model, 
and its Hamiltonian reads:
\begin{equation*}
H(\{\mathbf{\sigma}\})=\sum_{(i,j)}\epsilon_{ij}(\sigma_i, \sigma_j)
\text{\hspace{1cm} with }
\end{equation*}
\begin{align}\label{Eq:H}
 \epsilon_{ij}(\sigma_i, \sigma_j)=
\begin{cases}
 E_0 &\text{ if } (\sigma_i, \sigma_j)=(\sigma_i^*, \sigma_j^*)\\
 E_1 &\text{ otherwise}
\end{cases}
\end{align}
and
$(\sigma_i^*, \sigma_j^*)$ are randomly drawn among the $M\times M$ possible couples $(\sigma_i,\sigma_j)$ 
(independently for any couple of neighbors $(i,j)$).
For simplicity we will take $E_0=0$.
We believe that singling out one random couple of variables per link makes the 
model more frustrated than the usual Potts glass \cite{Potts} and the random-permutation versions of Ref. \cite{PermutationPotts,PermutationPotts2}. 
This is manifest in dimension $D=1$. For these models, after having chosen the value of the first Potts variable, 
one can easily find sequentially the configuration of the next variable that minimizes the energy, because for each value
of one variable, there exists a value of the neighboring one that can minimize the energy of the link.
For the Super-Potts glass, instead, there is only one particular configuration of both variables
that minimizes the energy of the link, and not all the links can be satisfied simultaneously even in $D=1$.
The Super-Potts glass can easily be generalized to more complicated choices of the link-energy, {\it e.g.}
$\epsilon_{ij}(\sigma_i, \sigma_j)$ randomly drawn from a Gaussian distribution. 
In this way, in the limit $M\rightarrow\infty$ one ends up with a random energy model on each link \cite{REM,FPR}.

\begin{table}
\begin{tabular}{|c|c|c|c|c|}
\hline
$M$ & $\beta_{RS}$ & $\beta_d$ & $\beta_k$ & $q_1(\beta_d)-q_0(\beta_d)$\\ \hline
4 & 2.0841(9) & 2.07(3) & 2.07(3) & 0 \\ \hline
10 & 1.9658(6) & 1.949(12) & 1.949(12) & 0\\ \hline
20 & 2.306(1) & 2.215(4) & 2.229(1) & 0.2623(1)\\ \hline
50 & 3.255(6) & 2.589(7) & 2.665(3) & 0.5772(7)\\ \hline
\end{tabular}
\caption{\label{Tab:Tc} $\beta_{RS}$, $\beta_d$, $\beta_k$ and the difference $q_1-q_0$ at the dynamical transition 
for different values of $M$ 
for the fully connected MF version of the Super-Potts model.}
\end{table}
We first present the analytical solution of the fully connected MF Super-Potts glass. 
The corresponding Hamiltonian is the one in eq. (\ref{Eq:H}) with the sum over all the pairs of Potts variables and the energy
that scales as $E_1=\frac{e_1}{\sqrt{N}}$, with $e_1=O(1)$ for finite $M$ and $N$ being the total number of Potts variables.
We sketch briefly the main steps of the computation and the results, 
more details can be found in the supplementary material.
The replica method allows one to compute the average free energy $f=\overline{f_{\epsilon}}$, 
where the bar indicates the average over the disorder,
in terms of the partition function of $n$ replicas:
\begin{equation}
 e^{-\beta Nnf}=\lim_{n\rightarrow0}\overline{Z^n}=
 \lim_{n\rightarrow0}\overline{\sum_{\{\boldsymbol{\sigma}\}}\prod_{i,j} e^{-\beta \sum_{a=1}^n \epsilon_{ij}(\sigma_i^a\sigma_j^a)}}.
\end{equation}
Repeating standard procedures \cite{Cavagna}, {\it i.e.} computing the average over the disorder, expanding the exponential for large $N$ 
and introducing Gaussian integrals over an auxiliary matrix $Q_{ab}$, we obtain:
\begin{equation}
 \overline{Z^n}\propto\sum_{\{\boldsymbol{\sigma}\}}\int\prod_{a<b}dQ_{ab}e^{-NA(\mathbf{Q},{\{\boldsymbol{\sigma}\}})}\propto\int d\mathbf{Q} e^{-NS(\mathbf{Q})} 
\label{Eq:Zn}
 \end{equation}
\begin{equation}
 \text{with\hspace{0.2cm}}A(\mathbf{Q},{\{\boldsymbol{\sigma}\}})=C\sum_{a<b}Q_{ab}^2-\frac{1}{N}\sum_{a<b}2C\sum_{i=1}^N\delta_{\sigma_i^a\sigma_i^b}Q_{ab}
\label{Eq:A(Q)}
\end{equation}
where we defined $C=(\frac{\beta e_1}{M})^2$.
We have chosen $e_1=M$ in order to reabsorb the scaling with $M$ of the critical temperature. The integral over $\mathbf{Q}$
is performed by the saddle-point method. The saddle point value of $Q_{ab}$, defined by the equation  
$\frac{dA(\mathbf{Q},{\{\boldsymbol{\sigma}\}})}{d\mathbf{Q}}=0$, corresponds to the average value 
of the overlap $\frac{1}{N}\sum_i\delta_{\sigma_i^a\sigma_i^b}$. By using the replica symmetric (RS) Ansatz, we
restrict the possible forms of $Q_{ab}$ to $Q_{ab}=q_{RS}$. Within this assumption the saddle-point equation
simplifies to:
$$q_{RS}=\int \prod_{\tau=1}^M\frac{dh_{\tau}}{\sqrt{4\pi}}e^{-\frac{h_{\tau}^2}{4}}
\frac{\sum_{\tau=1}^Me^{2\sqrt{Cq_{RS}}h_{\tau}}}{(\sum_{\tau=1}^Me^{\sqrt{Cq_{RS}}h_{\tau}})^2}.$$
Here and in the following, we shall solve these kinds of $M$-dimensional integrals by the Monte Carlo method.
Note that even when the RS solution is the correct, stable one, $q_{RS}$ is different from zero.
In order to analyze whether the RS solution is the correct one, we have also studied its local stability by diagonalizing the Hessian of the action:
$G_{ab,cd}=\left.\frac{d^2 S(Q_{ab})}{dQ_{ab}dQ_{cd}}\right|_{Q_{ab}=q_{RS}}$ \cite{dAT}.
One eigenvalue is always larger than 0, while the other one becomes negative at $T_{RS}(M)$,
indicating that the RS solution becomes unstable at low temperature. The values of $T_{RS}(M)$ are listed in Table \ref{Tab:Tc} for $M=4,10,20,50$.
Below $T_{RS}(M)$ one necessarily has to look for a RSB solution.
The next step is therefore to assume a 1RSB Ansatz \cite{1RSB} for the matrix $Q_{ab}$,
which is parametrized by three parameters $q_0$, $q_1$, $0\leq m\leq 1$.
We are interested in finding $T_d$, $T_K$ and deciding whether the transition is continuous or discontinuous; 
all this information can be obtained in the limit $m\rightarrow1$ \cite{Compl_Monasson}. 
In this case $q_0=q_{RS}$ and the saddle point equation on $q_1$ reads:
\begin{align*}
q_1&=\int\prod_{\tau=1}^M\frac{d\eta_{\tau}}{\sqrt{4\pi}}
\frac{e^{-\frac{\eta_{\tau}^2}{4}}}{\sum_{\tau=1}^Me^{C(q_1-q_{RS})+\sqrt{Cq_{RS}}\eta_{\tau}}}\times\\
&\times\int \prod_{\tau'=1}^M\frac{dh_{\tau'}}{\sqrt{4\pi}}e^{-\frac{h_{\tau'}^2}{4}}
\frac{\sum_{\tau'=1}^Me^{2(\sqrt{C(q_1-q_{RS})}h_{\tau'}+\sqrt{Cq_{RS}}\eta_{\tau})}}{\sum_{\tau'=1}^Me^{\sqrt{C(q_1-q_{RS})}h_{\tau'}+\sqrt{Cq_{RS}}\eta_{\tau}}}.
\end{align*}
Note that $q_1=q_{RS}$ is always a solution. As usual, we locate $T_d$ as the highest temperature at which one finds a solution $q_1\neq q_0$
and $T_K$ as the temperature at which the configurational entropy vanishes \cite{footnote}.
We found that for large values of $M$ ($M=20,50$) $q_1$ emerges discontinuously from $q_0$,
and $T_K(M)<T_d(M)$, signaling that the transition is 1-RSB, {\it i.e.} glass transition-like.
For smaller $M$ ($M=4,10$) instead, $q_1$ emerges continuously from $q_0$ and $T_K(M)=T_d(M)$,
meaning that the transition becomes continuous and similar to the one of MF spin-glasses in 
a field, {\it i.e.} of FRSB type.  
The difference between $q_0$ and $q_1$ at $T_d$ grows for larger $M$ indicating that increasing $M$ indeed favors 
structural glass-like behavior.
The values of $T_d(M)$, $T_K(M)$ and $q_1-q_0$ at $T_d$ are listed in Table \ref{Tab:Tc}.
In agreement with the previous results, for $M=4$ and $M=10$, 
the critical temperatures are compatible within the error with $T_{RS}$  \cite{footnote2}.


As discussed previously, three dimensional glass models may behave quite differently from their 
MF counterparts. It is therefore crucial to check that the Super-Potts glass still behaves
like a glass beyond MF. To this aim, we performed Monte Carlo (MC) numerical simulations of the model on 
a cubic lattice. We use the parallel tempering algorithm \cite{paralleltemp} 
to thermalize the system at low temperatures, running it simultaneously at 30 different
temperatures. Four replicas have been simulated in parallel,
letting them evolve independently 
with the same realization of disorder.
We measure the overlap $q$ between two of them, replicas $a,b$, as 
$q_{ab}=\frac{1}{N}\sum_{i=1}^N\delta_{\sigma_i^a, \sigma_i^b}.$
We check the equilibration dividing the first measurements into bins with a
logarithmically growing size, and we assume that the
system has reached the equilibrium when the probability distribution of the overlap $P(q)$
between the first two replicas is equal to $P(q)$ of the second two replicas
inside the last bin, and with respect to the precedent bin (practically we check the first four moments of $q$).
Equilibration time is of the order of $10^8$ MC steps for systems with $M=30$ and size $L=8$.
Once the system is thermalized, we run standard MC simulations
to measure dynamical correlation functions. 
Disorder averages were performed over 30 samples, while thermal ones over 100 trajectories.
\begin{figure}[t]	
\includegraphics[width=0.9\columnwidth]{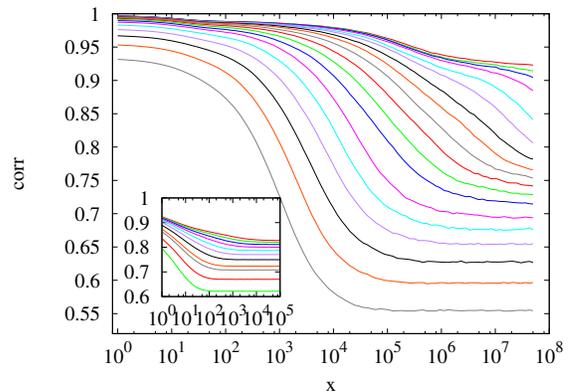}
\caption{Two-time correlation function for systems with $M=30$ and $L=8$ 
(main panel, inverse temperature $\beta$ equally spaced in $[0.28,0.85]$,  from left to right) and 
with $L=12$ $M=4$ (inset, $\beta$ equally spaced in $[0.76,1.09]$, from left to right).
}
\label{Fig:corr}
\end{figure}
\begin{figure}[t]	
\includegraphics[width=0.9\columnwidth]{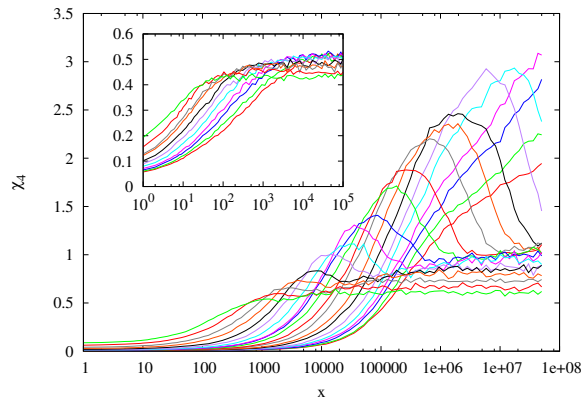}
\caption{Four point susceptibility for a system with 
$L=8$ $M=30$ (main panel) and with $L=12$ $M=4$ (inset). Temperatures as in Fig. \ref{Fig:corr}.}
\label{Fig:chi4}
\end{figure}
The behavior of the two times correlation (brackets indicate thermal average):
$$C(t)=\frac{1}{N}\overline{\sum_i\<\sigma_i(0)\sigma_i(t)\>}$$
is shown in Fig. \ref{Fig:corr} for $M=30$ \cite{footnote3}. By lowering the temperature the
two-steps relaxation characteristic of glass-forming liquids emerges 
(For $M=30$ the true plateau, corresponding in the peak of the susceptibility, 
is preceded by a first plateau that saturates at low enough temperature). 
Note that the asymptotic value of $C(t)$, $C(\infty)\equiv q_0$, is non zero since 
the Super-Potts glass, as many other disordered models introduced previously \cite{Potts, PermutationPotts, PermutationPotts2}, 
has no symmetry precluding $q_0$  from being different from zero (in consequence the two steps relaxation emerges on top of $q_0$) 
\cite{BinderKobbook}. The value of $q_0$ grows lowering the temperature,
as found also in the MF model, starting from $q_0=1/M$ at $T=\infty$.
For small values of $M$, instead, 
one finds a relaxation similar to the one of spin-glasses in a field, as shown in the inset
for $M=4$. In Fig. \ref{Fig:chi4} we show that the evolution of the four point susceptibility $\chi_4(t)$, defined as
$$
\frac{1}{N}\sum_{i,j}\((\overline{\<\sigma_i(0)\sigma_i(t)\sigma_j(0)\sigma_j(t)\>-\<\sigma_i(0)\sigma_i(t)\>\<\sigma_j(0)\sigma_j(t)\>}\)),
$$
confirms this trend: $\chi_4(t)$ is peaked, its maximum takes place at the time at which the correlation escape from the plateau 
and grows when lowering the temperature as it happens for super-cooled liquids. This behavior, present for $M=30$, 
is markedly different from the one shown in the inset for $M=4$. 
For $M$s in between the two presented values
the system actually seems to show a mixed behaviour, for instance $\chi_4(t)$  
shows a peak but also a growing plateau.
\begin{figure}
\begin{center}
\includegraphics[width=0.35\textwidth]{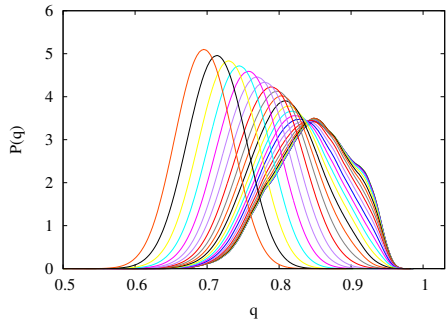}
 \caption{$P(q)$ for a system with $L=10$ $M=20$. $\beta$ equally spaced in $[0.55, 1.3]$ (from left to right).}
\label{Fig:L10M20}
\end{center}
\end{figure}
We also studied the overlap distribution $P(q)$. 
Although of course one would need much larger sizes to provide convincing evidences of a phase transition, 
our results shown in Fig. \ref{Fig:L10M20} suggest that if there is a transition then it should 
be discontinuous already for $M=20$, since a second peak seems to appear discontinuously 
at small temperatures as if a 1RSB transition were indeed taking place.
Overall our  numerical results indicate that at large 
$M$ ($M\gtrsim 20$) the Super-Potts glass behaves similarly to glass-forming liquids whereas for smaller $M$s 
analogously to a spin-glass in a field, in agreement with the MF treatment presented before.  


In conclusion we introduced a new model, the Super-Potts glass, and showed 
that is the first long-sought example of glassy disordered system with 
pair-wise interactions, solved by a 1RSB Ansatz at the MF level, and which has in three dimensions
a phenomenological behaviour strongly reminiscent of glass-forming liquids. 
In particular it shows stretching (non-exponential behaviour) and two steps relaxation for the correlation function, 
a time for the relaxation from the plateau that seems to diverge at finite temperature, a growing peak in the
four point correlation function and a discontinuous peak appearing in the $P(q)$. 
The glassy behaviour is only found for sufficiently high numbers $M$ of values that the Potts variables can take.
This is reasonable if we think to a real-world structural glass, where the degrees of freedom, i.e. 
the position of particles, can take infinite values. 
Compared to previous models for which the glassy behaviour does not survive in finite dimensions, 
the Super-Potts glass is more frustrated and this enhances its stability. Indeed, we computed 
the so called surface energy cost, $Y$, to disrupt amorphous order as done in \cite{NO1RSBbethe} 
and found a value of $Y/T_K$  which is an order of magnitude higher than in previous models for large values of $M$, 
e.g. $M=50$. 
There are several extensions of our work worth pursuing further. First, it would be interesting to 
clarify how the transition between the glass-like to the spin-glass like behaviour induced 
by decreasing the value of $M$ takes place, both in mean-field and in finite dimensions. 
A possible scenario, inspired by the behaviour of the 2+4 spin MF model,
 is the following \cite{1FRSB}: Whereas at small $M$ there is a pure FRSB phase and at large $M$ a pure 1RSB phase, 
 at intermediate $M$, by decreasing the temperature, there is first a RS to FRSB transition,
and then, lowering the temperature further, there is a transition to a  1+FRSB 
in which $P(q)$ has a continuous part but also develops a discontinuous
peak. This is consistent with the fact that for intermediate values of M the correlation function and 
the four point susceptibility show mixed features characteristic both of the 1RSB and FRSB phases. 
Another research direction for future studies is solving exactly the Super-Potts model on Bethe lattices.  
This would provide a good approximation to the 3D case since, as we found in numerical simulations, 
the behaviour on cubic and Bethe lattices with connectivity $C=6$ is qualitatively and also quantitatively
similar. The exact solution of models on the Bethe lattices can be obtained via the cavity method which 
in the case of the Super-Potts glass, however, is particularly challenging \cite{footnote4}.
It could be also interesting to apply the trick used in Ref. \cite{PermutationPotts} to obtained
a modified version of the model that could have a symmetric $P(q)$, allowing an easier thermalization and
more extensive numerical simulations.
Finally, another interesting route to follow in order to clarify the behaviour of the 3D model is performing 
a renormalization group analysis. Since the model has  pair-wise interactions, this can be naturally done
by the Migdal-Kadanoff approximation.

\begin{acknowledgments}
We acknowledge support from the ERC grants NPRGGLASS. We thank F. Caltagirone, U. Ferrari, M. Moore, M. Muller, 
F. Ricci-Tersenghi and M. Tarzia,  for useful discussions.   
\end{acknowledgments}

\section{Supplementary Material}

In the following we provide more details on the MF solution presented in the main text.\\ 
In order to compute the replicated free energy:
\begin{equation*}
 \overline{Z^n}=
\overline{\sum_{\{\boldsymbol{\sigma}\}}\prod_{i,j} e^{-\beta \sum_{a=1}^n \epsilon_{ij}(\sigma_i^a\sigma_j^a)}}=
\end{equation*}
we introduce $(\overline{\gamma},\overline{\tau})$ as the couple for which $\epsilon_{ij}=E_0$,
and making explicit the average over the disorder, that is the average over the randomly chosen $(\overline{\gamma},\overline{\tau})$,
we obtain:
\begin{equation*}
 \overline{Z^n}=
\sum_{\boldsymbol{\sigma}}\prod_{i,j}\frac{1}{M^2}\sum_{\overline{\gamma},\overline{\tau}} 
e^{-\beta E_1\sum_{(\gamma, \tau)\neq(\overline{\gamma},\overline{\tau})}\sum_{a=1}^n \delta_{(\sigma_i^a,\sigma_j^a),(\gamma,\tau)}}
\end{equation*}

Expanding around small energies $E_1=\frac{e_1}{\sqrt{N}}$ and reexponentiating, the expression becomes:

\begin{equation*}
 \overline{Z^n}=
\sum_{\boldsymbol{\sigma}}\prod_{i,j} 
e^{-\frac{\beta^2 E_1^2}{2M^2}\sum_{a,b=1}^n \((\sum_i\delta_{\sigma_i^a,\sigma_i^b}\))^2}.
\end{equation*}

Performing the usual Hubbard-Stratonovich transformation to eliminate the quadratic term, 
with the help of an auxiliary matrix $Q_{ab}$, we obtain:

$$\overline{Z^n}\propto\sum_{\boldsymbol{\sigma}}\int\prod_{a<b}dQ_{ab}e^{-NA(Q)}$$
with
$$A(Q)=C\sum_{a<b}Q_{ab}^2-\frac{1}{N}\sum_{a<b}2C\sum_{i=1}^N\delta_{\sigma_i^a\sigma_i^b}Q_{ab}$$
and $C=\frac{(\beta e_1)^2}{M^2}$. This is the equation quoted in the main text.
\subsection{RS ansatz}
By using the replica symmetric ansatz $Q_{ab}=q$, we obtain
$$\overline{Z^n}\propto\int dq e^{-NS(q)}$$
with 

\begin{align*}
 S(q)=&Cq^2\frac{n^2-n}{2}+Cnq+\\
&-n\int \prod_{\tau=1}^M\frac{dh_{\tau}}{\sqrt{4\pi}}e^{-\frac{h_{\tau}^2}{4}}
\log(\sum_{\tau=1}^Me^{\sqrt{Cq}h_{\tau}})
\end{align*}

The overlap $q_{RS}$ satisfies the self-consistent equation obtained imposing $\frac{dS(q)}{dq}=0$ in the limit $n\rightarrow0$:

$$q_{RS}=\int \prod_{\tau=1}^M\frac{dh_{\tau}}{\sqrt{4\pi}}e^{-\frac{h_{\tau}^2}{4}}
\frac{\sum_{\tau=1}^Me^{2\sqrt{Cq_{RS}}h_{\tau}}}{(\sum_{\tau=1}^Me^{\sqrt{Cq_{RS}}h_{\tau}})^2}$$

\subsection{RS stability}

To study the RS stability we look at the Hessian of $S(Q_{ab})$.
The second derivatives are:

\begin{align*}
G_{abcd}=&\frac{d^2 S(Q_{ab})}{dQ_{ab}dQ_{cd}}=\\
=&2C\delta_{ab,cd}-(2C)^2(\<\delta_{\sigma_a,\sigma_b}\delta_{\sigma_c,\sigma_d}\>-
\<\delta_{\sigma_a,\sigma_b}\>\<\delta_{\sigma_c,\sigma_d}\>)
\end{align*}
In particular
$$G_{abab}=2C-(2C)^2(\<\delta_{\sigma_a,\sigma_b}\>-
\<\delta_{\sigma_a,\sigma_b}\>^2)=P$$

$$G_{abac}=-(2C)^2(\<\delta_{\sigma_a,\sigma_b,\sigma_c}\>-
\<\delta_{\sigma_a,\sigma_b}\>\<\delta_{\sigma_a,\sigma_c}\>)=Q$$

$$G_{abcd}=-(2C)^2(\<\delta_{\sigma_a,\sigma_b}\delta_{\sigma_c,\sigma_d}\>-
\<\delta_{\sigma_a,\sigma_b}\>\<\delta_{\sigma_c,\sigma_d}\>)=R$$

$$\<\delta_{\sigma_a,\sigma_b}\>=q_{RS}=\int \prod_{\tau=1}^M\frac{dh_{\tau}}{\sqrt{4\pi}}e^{-\frac{h_{\tau}^2}{4}}
\frac{\sum_{\tau=1}^Me^{2\sqrt{Cq_{RS}}h_{\tau}}}{(\sum_{\tau=1}^Me^{\sqrt{Cq_{RS}}h_{\tau}})^2}$$

$$\<\delta_{\sigma_a,\sigma_b,\sigma_c}\>=\int \prod_{\tau=1}^M\frac{dh_{\tau}}{\sqrt{4\pi}}e^{-\frac{h_{\tau}^2}{4}}
\frac{\sum_{\tau=1}^Me^{3\sqrt{Cq_{RS}}h_{\tau}}}{(\sum_{\tau=1}^Me^{\sqrt{Cq_{RS}}h_{\tau}})^3}$$

$$\<\delta_{\sigma_a,\sigma_b}\delta_{\sigma_c,\sigma_d}\>=\int \prod_{\tau=1}^M\frac{dh_{\tau}}{\sqrt{4\pi}}e^{-\frac{h_{\tau}^2}{4}}
\frac{\((\sum_{\tau=1}^Me^{2\sqrt{Cq_{RS}}h_{\tau}}\))^2}{(\sum_{\tau=1}^Me^{\sqrt{Cq_{RS}}h_{\tau}})^4}$$

The eigenvalues are
$$\lambda_1=2C-(2C)^2\((\<\delta_{\sigma_a,\sigma_b}\>-4\<\delta_{\sigma_a,\sigma_b,\sigma_c}\>
+3\<\delta_{\sigma_a,\sigma_b}\delta_{\sigma_c,\sigma_d}\>\))$$

$$\lambda_2=2C-(2C)^2\((\<\delta_{\sigma_a,\sigma_b}\>-2\<\delta_{\sigma_a,\sigma_b,\sigma_c}\>
+\<\delta_{\sigma_a,\sigma_b}\delta_{\sigma_c,\sigma_d}\>\))$$

The first one is always larger than 0, while $\lambda_2$ becomes negative at $T_{RS}(M)$. 

\subsection{1RSB ansatz}

In the 1RSB ansatz we have three parameters $q_0$, $q_1$, $m$ and 

\begin{align*}
S(Q_{ab})=&\frac{C}{2}n(q_1^2(m-1)+q_0^2(n-m))+nCq_1+\\ 
&-\frac{n}{m}
\int\prod_{\tau=1}^M\frac{d\eta_{\tau}}{\sqrt{4\pi}}e^{-\frac{\eta_{\tau}^2}{4}}
\log\[[\int \prod_{\tau'=1}^M\frac{dh_{\tau'}}{\sqrt{4\pi}}e^{-\frac{h_{\tau'}^2}{4}}\times \right.\\
&\left. \times \((\sum_{\tau'=1}^Me^{\sqrt{C(q_1-q_0)}h_{\tau'}+\sqrt{Cq_0}\eta_{\tau}}\))^m\]].
\end{align*}

We are interested in obtaining $T_c$ and finding whether the transition is continuous or discontinuous. In consequence 
we have just to focus on $m\rightarrow 1$. In this case $q_0=q_{RS}$ and we can find $q_1$ self-consistently imposing
$\frac{dS(q_1,q_{RS})}{dq_1}=0$. We can expand the resulting equation around $m=1$, and at the 1st order we find 
(the 0th order is 0):

\begin{align*}
q_1&=f(q_1,q_{RS},\beta)=\int\prod_{\tau=1}^M\frac{d\eta_{\tau}}{\sqrt{4\pi}}
\frac{e^{-\frac{\eta_{\tau}^2}{4}}}{\sum_{\tau=1}^Me^{C(q_1-q_{RS})+\sqrt{Cq_{RS}}\eta_{\tau}}}\times\\
&\times\int \prod_{\tau'=1}^M\frac{dh_{\tau'}}{\sqrt{4\pi}}e^{-\frac{h_{\tau'}^2}{4}}
\frac{\sum_{\tau'=1}^Me^{2(\sqrt{C(q_1-q_{RS})}h_{\tau'}+\sqrt{Cq_{RS}}\eta_{\tau})}}{\sum_{\tau'=1}^Me^{\sqrt{C(q_1-q_{RS})}h_{\tau'}+\sqrt{Cq_{RS}}\eta_{\tau}}}
\end{align*}
that is valid for $\beta\in\[[\beta_d,\beta_K\]]$.


\begin{thebibliography}{100}
\bibitem{footnote}
The complexity is computed as the derivative of the free energy with respect to $m$:
$\Sigma(m)=m^2\frac{\partial (\beta f(q_0,q_1,m))}{\partial m}$ \cite{Compl_Monasson}.
\bibitem{footnote3} $C(t=1)$ is, for small $\beta$, 
quite different from 1 because of the degeneration in the configurations of the spins 
that do not satisfy any of the couplings: they can assume at least $M-6$ different values without a change in the energy.
\bibitem{footnote4}
Because of the quenched disorder, each spin is different from the others; for this reason
already at the RS level, the probability analysis of the distribution of the cavity messages 
have a non trivial population of messages. Moreover in this model $q_0\neq0$, 
leading to other complications.
\end{thebibliography}
\end{document}